\documentstyle{mn}
\input epsf.tex
\def\figinsert#1#2{\epsfbox{#1} \message{#2} }    
\def \Mpc {{\,{\it h}^{-1} {\rm Mpc}}}

\def \lsim {\mathrel{\rlap{\lower 3pt\hbox{$\mathchar"218$}}
       \raise 2.0pt\hbox{$\mathchar"13C$}}}
\def \gsim {\mathrel{\rlap{\lower 3pt\hbox{$\mathchar"218$}}
       \raise 2.0pt\hbox{$\mathchar"13E$}}}
\def \xcc{\xi_{\rm cc}}
\def\etal{\it et al. \rm}

\def\hmpc{{\it h}$^{-1} {\rm Mpc}\,$}

\def\gsim{ \lower .75ex \hbox{$\sim$} \llap{\raise .27ex \hbox{$>$}} }
\def\lsim{ \lower .75ex \hbox{$\sim$} \llap{\raise .27ex \hbox{$<$}} }
\def\spose#1{\hbox to 0pt{#1\hss}}
\def\simlt{\mathrel{\spose{\lower 3pt\hbox{$\mathchar"218$}}
     \raise 2.0pt\hbox{$\mathchar"13C$}}}
\def\simgt{\mathrel{\spose{\lower 3pt\hbox{$\mathchar"218$}}
     \raise 2.0pt\hbox{$\mathchar"13E$}}}
\def\gsim{ \lower .75ex \hbox{$\sim$} \llap{\raise .27ex \hbox{$>$}} }
\def\lsim{ \lower .75ex \hbox{$\sim$} \llap{\raise .27ex \hbox{$<$}} }
\def\simprop{ \lower .75ex \hbox{$\sim$} \llap{\raise .27ex \hbox{$\propto$}} }
\title[Cluster Clustering]
      {Cluster correlation functions in N-body simulations}
\author[V.R.Eke, S.Cole, C.S.Frenk and J.F.Navarro]
       {Vincent R. Eke$^{1,3}$, Shaun Cole$^{1,4}$, Carlos S. Frenk$^{1,5}$ and
Julio F. Navarro$^{1,2,6}$ \\
$^{1}$Department of Physics, University of Durham, Science
Laboratories, South Rd, Durham DH1 3LE\\
$^{2}$Steward Observatory, The University of Arizona, Tucson,
AZ 85721, U.S.A. \\
$^{3}$V.R.Eke@Durham.ac.uk\\
$^{4}$Shaun.Cole@Durham.ac.uk\\
$^{5}$C.S.Frenk@Durham.ac.uk\\
$^{6}$jnavarro@as.arizona.edu }
\begin{document}

\maketitle

\begin{abstract}

The correlation function of galaxy clusters has frequently been used as a
test of cosmological models. A number of assumptions are implicit in the
comparison of theoretical expectations to data. Here we use an ensemble of
ten large N-body simulations of the standard cold dark matter cosmology to
investigate how cluster selection criteria and other uncertain factors
influence the cluster correlation function. Our study is restricted to the
idealized case where clusters are identified in the three dimensional mass
distribution of the simulations. We consider the effects of varying the
definition of a cluster, the mean number density (or equivalently the
threshold richness or luminosity) in a catalogue, and the assumed
normalization of the cosmological model; we also examine the importance of
redshift space distortions. We implement five different group-finding
algorithms and construct cluster catalogues defined by mass, velocity
dispersion or a measure of X-ray luminosity. We find that different
cluster catalogues yield correlation functions which can differ from one
another by substantially more than the statistical errors in any one
determination. For example, at a fixed
number density of clusters, the characteristic clustering length can vary
by up to a factor of $\sim 1.5$, depending on the precise procedure
employed to identify and select clusters. For a given cluster selection
criteria, the correlation length typically varies by $\sim 20\%$ in catalogues
spanning the range of intercluster separations covered by the APM and
Abell (richness class $\gsim 1$) catalogues. Distortions produced by
peculiar velocities in redshift space enhance the correlation function at
large separations and lead to a larger clustering length in redshift space
than in real space. The sensitivity of the cluster correlation function to
various uncertain model assumptions substantially weakens previous
conclusions based on the comparison of model predictions with real data.
For example, some of our standard cold dark matter cluster catalogues
agree better with published cluster clustering data (particularly on small
and intermediate scales) than catalogues constructed from similar
simulations by Bahcall \& Cen and Croft \& Efstathiou.  Detailed modelling
of cluster selection procedures including, for example, the effects of
selecting from projected {\it galaxy} catalogues is required before the
cluster correlation function can be regarded as a high precision constraint
on cosmological models.

\end{abstract}

\begin{keywords}
cosmology:theory -- large-scale structure of Universe
\end{keywords}

\section{Introduction}

The clustering strength of rich galaxy clusters has long been used as a
constraint on models of large scale structure. The two-point
correlation function, $\xcc$, was first estimated for a sample of about 100
rich Abell clusters by Bahcall \& Soneira (1983) and by Klypin \& Kopylov
(1983) who noted that clusters have a larger clustering amplitude than
galaxies. This difference has a natural explanation in theories in which
large scale structure grows by gravitational amplification of small
fluctuations in an initially Gaussian density field. In such theories,
collapsed objects form near peaks of the initial density field and a
clustering pattern which depends on the height of the peak is imprinted at
the epoch of formation (Kaiser 1984, Barnes \etal 1985).

Although the statistics of rare peaks provide an appealing explanation for
the different clustering strength of galaxies and clusters, it soon became
apparent that the early estimates of $\xcc$ were inconsistent with the
predictions of the cold dark matter model, the paradigm of gravitational
clustering theories (Davis \etal 1985). Using N-body simulations of this
model, White \etal (1987) calculated a cluster clustering length,
$r_0\simeq 11$\hmpc ($r_0$ is defined as the separation at which $\xcc=1$), 
whereas 
Bahcall \& Soneira (1983) had obtained $r_0 \sim 25$\hmpc for Abell
clusters of richness class $R\ge 1$. Bahcall \& Soneira's estimate helped
to motivate an alternative explanation for the formation of structure, based
on fluctuations seeded by cosmic strings (Turok 1983, Turok \&
Brandenberger 1986).

Sutherland (1988) pointed out that the apparent clustering amplitude of
rich clusters would be artificially enhanced if intrinsically poor clusters
which happened to lie near the line-of-sight to a rich cluster -- and thus
appear rich in projection -- were included in a rich cluster sample. A
signature of this effect is an anisotropy in the correlation function which
appears stronger along the line-of-sight than in the perpendicular
direction.  Soltan (1988) and Sutherland \& Efstathiou (1991) showed that
such anisotropies were clearly present in Bahcall \& Soneira's sample and
they, as well as Dekel \etal (1989), argued that correcting for this effect
would lower the clustering length of rich Abell clusters to $\sim
14$\hmpc.
Nevertheless, Postman, Huchra \& Geller (1992), using a sample of 351 Abell
clusters of richness class $R\ge 0$, supported Bahcall \& Soneira's
original estimate.

Further progress had to await the construction of new cluster
catalogues. These finally began to arrive in the early 1990s. Dalton \etal
(1992) and Lumsden \etal (1992) constructed the first automated cluster
catalogues using the positions and magnitudes of galaxies in photographic
plates scanned with the APM and Cosmos machines respectively; Lahav \etal
(1993), Romer \etal (1994) 
and Nichol, Briel \& Henry (1994) constructed cluster
catalogues using a combination of X-ray and optical data. The ensuing redshift
surveys led to new determinations of $\xcc$. The Oxford group
estimated $r_0=12.9 \pm 1.4$\hmpc (Dalton \etal 1992) from a sample of
about 200 APM clusters, and $r_0=14.3 \pm 1.2$\hmpc from an extended sample
of 364 APM clusters (Dalton \etal 1994); Romer \etal obtained $r_0=13.7 \pm
2.3$\hmpc from an X-ray flux-limited sample of 128 clusters.

Unfortunately the new cluster samples have not fully resolved the debate
surrounding $\xcc$. Bahcall \& Soneira (1983) argued that the measured
values of $r_0$ depend very strongly on cluster richness.
Bahcall \& West (1992) interpreted the
discrepancies between different samples as a reflection of their different
mean cluster richness, rather than as a result of contamination in Abell's
catalogue (see also Peacock \& West 1992). N-body simulations by Bahcall 
\& Cen (1992) seem to support this view, whereas simulations by 
Croft \& Efstathiou (1994) suggest that the dependence of $r_0$ on
cluster richness is weak. Furthermore, the Oxford group have claimed 
that the low values of $r_0$ that they obtain for APM cluster catalogues,
although much smaller than Bahcall \& Soneira's (1983) value are still 
inconsistent with the standard CDM cosmology and favour either CDM models
with a low mean density and a non-zero cosmological constant or mixed dark
matter models (Dalton \etal 1992, 1994, Croft \& Efstathiou 1994). They
base this conclusion on Croft \& Efstathiou's (1994) set of large N-body
simulations which, they argue, enable theoretical predictions for $\xcc$
to be made with better than $10\%$ accuracy over a wide range of scales.

The work of Bahcall \& Cen (1992), Croft \& Efstathiou (1994) and
Watanabe \etal (1994) has
highlighted how, as the observational data on $\xcc$ improves, the need
for precise theoretical predictions becomes increasingly important. Making
theoretical predictions which are relevant to the interpretation of the
data, however, is not straightforward, even for well specified models such
as CDM and its variants. In these models the evolution of the {\it mass}
density field on the relevant scales can indeed be predicted quite
accurately, particularly through large N-body simulations. The primary
difficulty lies in the uncertain identification of clusters in the models
with the real clusters found in galaxy surveys. 
Bahcall \& Cen, Croft \& Efstathiou and Watanabe \etal, like White 
\etal (1987), identified ``galaxy
clusters'' in their simulations with large mass concentrations in the
three-dimensional mass distribution. This is, of course, a very different
procedure from that applied to real data where clusters are identified
from the {\it projected } galaxy distribution using relatively complex
algorithms. Possible biases in statistics such as $\xcc$ which might be
introduced by this procedure remain largely unexplored.

In this paper, we address the restricted question of how the clustering
strength of model clusters identified using the full three-dimensional mass
distribution in N-body simulations depends on the details of the cluster
finding algorithm and on the way in which the cluster catalogues are 
constructed. For
definitiveness, we consider only one specific cosmological model, the
standard CDM model. We find that even in this idealized case, a wide range
of values of $r_0$ is obtained from samples selected and analyzed in ways
which are in principle equally valid approximations to the real situation. We
therefore conclude that much more detailed modelling is required before the
present data can be used as a high precision test of currently popular 
cosmological models.

In the following section we describe our simulations and methods for
constructing cluster catalogues. In Section 3 we present estimates of
the correlation function for these catalogues. In Section~4 we compare our 
results with those obtained in previous related studies. We discuss our 
findings and summarize our conclusions in Section~5. 

\section{Techniques}

\subsection{Numerical Simulations}

Our analysis is based on an ensemble of 10 CDM simulations performed with
the adaptive AP$^{3}$M code of Couchman (1991). Each simulation represents
a comoving periodic box of length $l_{\rm box}=256 \Mpc $
\footnote{Throughout this paper, we write Hubble's constant as $H_{0}=100h 
{\rm km s}^{-1} {\rm Mpc}^{-1}$}, containing $128^{3}$ particles, each of
mass $\sim2.2\times 10^{12}h^{-1}M_{\odot}$. The force softening (for an
equivalent Plummer potential) was chosen to be $\sim65h^{-1}$kpc and
remained fixed in comoving coordinates.

Initial conditions were laid down by perturbing particles from a uniform grid
using the Zel'dovich approximation (Efstathiou \etal 1985), and assuming the
BBKS CDM transfer function for the case of zero baryons and $H_{0} = 50{\rm
kms}^{-1}{\rm Mpc}^{-1}$ (Bardeen {\it et al} 1986). We defined the
expansion factor $a$ such that $\sigma_{8}=a$, where $\sigma_{8}$ is the
linear { \it rms} amplitude of mass fluctuations in spheres of radius $8
\Mpc$. Each simulation was evolved from $a = 0.05$ to $a = 0.63$ using a
timestep $\Delta a = 0.002$ and each took approximately two days of CPU on a
Decstation 5000/240. The timestep and initial redshift were chosen after
running a series of trial simulations in which these parameters were
varied. Although the cluster correlation function turned out to be
insensitive to these
variations, we found that adopting either a later starting time or a
larger timestep made a significant change to the abundance of clusters as
a function of both mass and temperature, while adopting earlier starting
times and smaller timesteps left them essentially unchanged.

In the following analysis we identify the present day with the epoch at
which $\sigma_{8}=0.5$ or $0.63$. These values were chosen to span the
range of normalizations that are consistent with the mass and abundance of
rich clusters of galaxies (White, Efstathiou \& Frenk 1993) and are lower
than would be required for consistency with the {\it COBE} microwave
background anisotropy measurements in the absence of a tensor mode
contribution to the anisotropy (Smoot \etal 1992, Wright \etal 1994).
Our simulations have a slightly smaller volume, but better mass and spatial
resolution than those of Croft \& Efstathiou (1994).

\subsection{Group-Finders}

In order to investigate the effect of varying the criteria by which
clusters are defined and selected, we used five different group-finding
algorithms to identify clusters of particles in the N-body simulations. In
all cases we considered only groups with 8 or more particles, corresponding
to a mass $M\ge 3.5\times 10^{13} h^{-1}M_{\odot}$.

The first was the standard friends-of-friends algorithm (Davis {\it et
al}. 1985) that links together particles closer than some specified
separation. We adopted a linking length of $b = 0.2$ times the
mean interparticle separation for which groups have typical mean
overdensities of $\sim 200$. We will refer to these groups as FOF.

Our second group-finder was the ``spherical overdensities"
algorithm described in more detail in Lacey \& Cole (1994). This algorithm
estimates the local density at the position of each particle by finding
the distance to its tenth nearest neighbour. Particles are ranked by local
density and, starting with the particle in the densest environment, they
are used as centres about which spheres are inflated until the mean
enclosed overdensity falls below a specfied threshold. Overlapping groups
are merged and centres recalculated until they lie within $0.2 \Mpc$ of
the centre-of-mass of the grouped particles. We adopted an overdensity
threshold of $180$. We shall refer to the resulting groups as SO.

Our third algorithm is a variant of the spherical
overdensities algorithm. Each particle is assigned an ``X-ray luminosity",
$L_{{\rm x},i}=\rho_{i}V_i$, where $\rho_{i}$ and $V_i$ denote
estimates of the local density and velocity dispersion obtained from the
ten nearest neighbours. The motivation for this choice is that, when
summed over a group of particles, the total ``luminosity" will depend on
density and temperature in the same way as bremsstrahlung emission, namely
$L_{\rm x}\propto\int \rho^{2}T_{\rm x}^{0.5}d^3r$. We calculate the mean 
X-ray luminosity density by summing all of the individual particle values 
and dividing by the volume of the simulation box.
Then the particles are ranked by X-ray
luminosity and spheres inflated until the luminosity contrast in the
spherical volume, $\delta L_{\rm x}/L_{\rm x}=10^{4}$. 
Although our simulations lack the spatial resolution
to define the X-ray emitting regions well, this group-finder does give a
higher weighting to the centres of the clusters and therefore has an
effect which is qualitatively similar to that which we seek to represent.
We will refer to groups identified in this way as SOX.

The final two group-finders we used are identical to those adopted by Croft \&
Efstathiou (1994) and allow us to make a direct comparison with their
results. The algorithm first locates potential group centres using the
FOF algorithm with $b=0.1$ to find tight knots of particles. Spheres of
radius either $0.5$ or $1.5 \Mpc$ are constructed about these centres and,
after merging overlapping spheres and recentring, the particles within
this radius are considered as a group. We denote these as CE(0.5) and
CE(1.5) respectively.

In summary, we have chosen to investigate five different group-finding
algorithms. The first, FOF, is the standard group-finder used
extensively in previous analyses of N-body simulations. This is a simple
and elegant algorithm which picks out most of the clusters identified as
such by eye, although occasionally it classes two distinct lumps as a
single group if they are linked by a tenuous bridge of particles.
Our second algorithm, SO, avoids this situation by working with the
average densities within spheres. The shortcoming of this method is
that it unnaturally forces the boundaries of groups to be spheres. Our
third algorithm, SOX, attempts to emulate selecting clusters by
X-ray emission. Given the nature and resolution of our simulations it is
only approximate, but it does succeed in giving increased weight to groups
containing dense cores. The fourth and fifth group-finders, CE(0.5) and
CE(1.5), we use for comparison with Croft \& Efstathiou (1994) who adopted
them as idealized 3-D versions of the galaxy counting algorithms that
were used in constructing the APM and Abell galaxy cluster catalogues
respectively. 


\begin{figure}
\centering
\centerline{\epsfysize=9.5truecm \figinsert{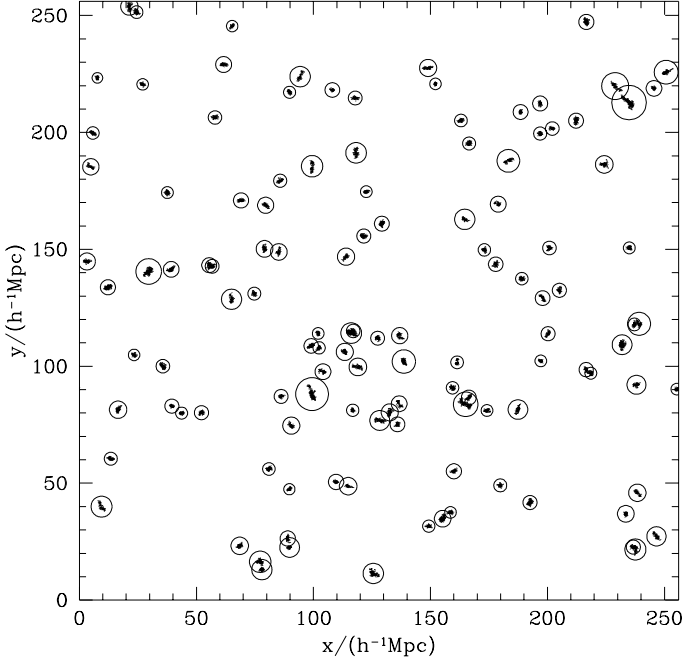}{441.0pt}}
\caption{A $50 \Mpc$ thick slice through one of the ten simulations with
$\sigma_8=0.63$. The clusters were selected using the friends-of-friends
(FOF) group-finder and adopting a lower mass cutoff such that the mean
cluster separation $d_{\rm c}=30 \Mpc$. Particles belonging to clusters are
shown as dots and each cluster is delineated by a circle around its centre
with radius equal to 1.5 times the true distance to the most distant
cluster particle.}
\end{figure}

\begin{figure}
\centering
\centerline{\epsfysize=9.5truecm \figinsert{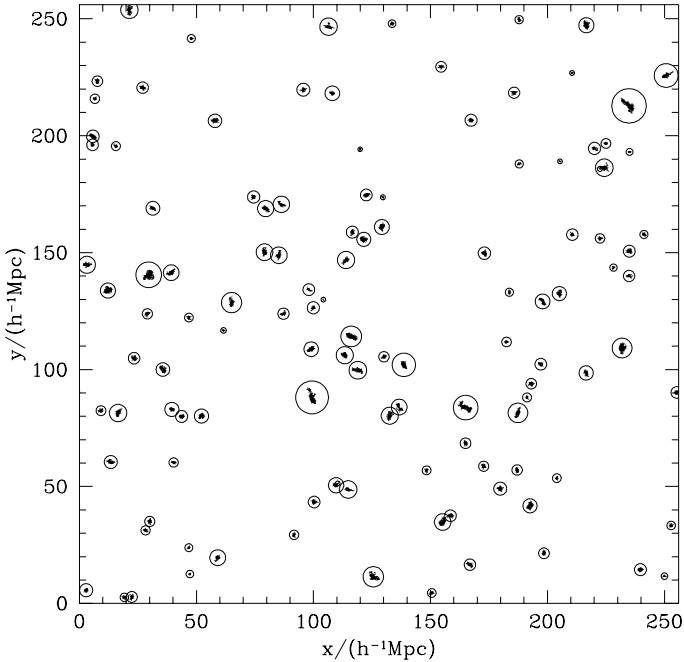}{441.0pt}}
\caption{As Fig. 1, but with X-ray luminosity rather than mass as the
selection statistic for the FOF groups.}
\end{figure}

\subsection{Cluster Selection}

The next step is to construct cluster catalogues from each set of groups
identified by our different algorithms. In the real world, the abundance
of clusters in a sample is determined by setting a threshold in apparent
optical richness or X-ray flux. In our models, for each choice of group-finder,
we ranked the groups identified in all $N_{\rm sim}=10$ simulations according
to mass ($M$), velocity dispersion ($v$), or X-ray luminosity
($L_{\rm x}$). We then selected the $N_{\rm clus}$ highest ranked
clusters to produce a cluster catalogue in each simulation for which the
mean intercluster separation is given by
\begin{equation}
 d_{\rm c}=(N_{\rm sim}/N_{\rm clus})^{1/3}l_{\rm box}.
\end{equation}
We investigate how the resulting cluster correlation function depends
on abundance, as parameterized by $d_{\rm c}$, as well as on the other
details of the cluster selection process.

Cluster X-ray luminosities were defined in one of two ways. For the SOX
groups, we simply summed the luminosities assigned to the individual
particles in the group. For clusters obtained with other group-finders we
first estimated a velocity dispersion (from the measured mass assuming an
isothermal density distribution), converted this into an X-ray temperature,
$T_{\rm x}$, by assuming that the specific kinetic energy in the dark matter
is equal to the specific thermal energy in the gas (Evrard 1990; Navarro, 
Frenk \& White 1995), and inferred an X-ray
luminosity from the empirical $L_{\rm x}-T_{\rm x}$ relation. For the last
step we adopted the mean relation given by David \etal (1993), $L_{\rm x}
\propto T_{\rm x}^{3.4}$. (Since we are only interested in 
ranking the clusters by X-ray luminosity, the constant of proportionality is 
immaterial.) We then assumed
a Gaussian scatter with variance varying linearly with ${\rm log_{10}}T_{\rm
x}$. This gave $\sigma_{{\rm log_{10}}L_{\rm x}}=0.58$ at 
${\rm log_{10}}T_{\rm x}=0.3$, 
and $\sigma_{{\rm log_{10}}L_{\rm x}}=0.31$ at ${\rm log_{10}}T_{\rm x}=0.9$.
This procedure essentially scrambles up a $v$-selected
catalogue by introducing some lower velocity dispersion clusters at the
expense of higher velocity dispersion ones.

For each of our group-finders and selection statistic, catalogues with
$20\leq d_{\rm c}\leq 70 \Mpc$ were created for both $\sigma_{8}=0.5$ and
$\sigma_{8}=0.63$. Figs.~1 and~2 show a slice through one of the
simulations. In both cases, clusters were found using FOF, but the
clusters in Fig.~1 were selected by mass whereas those in Fig.~2 were
selected by ``X-ray" luminosity. The ``X-ray" sample can be seen to have some
very small clusters that have come in at the expense of more extended
objects in the mass selected sample. 

\begin{figure*}
\centering
\centerline{\epsfysize=15.5truecm \figinsert{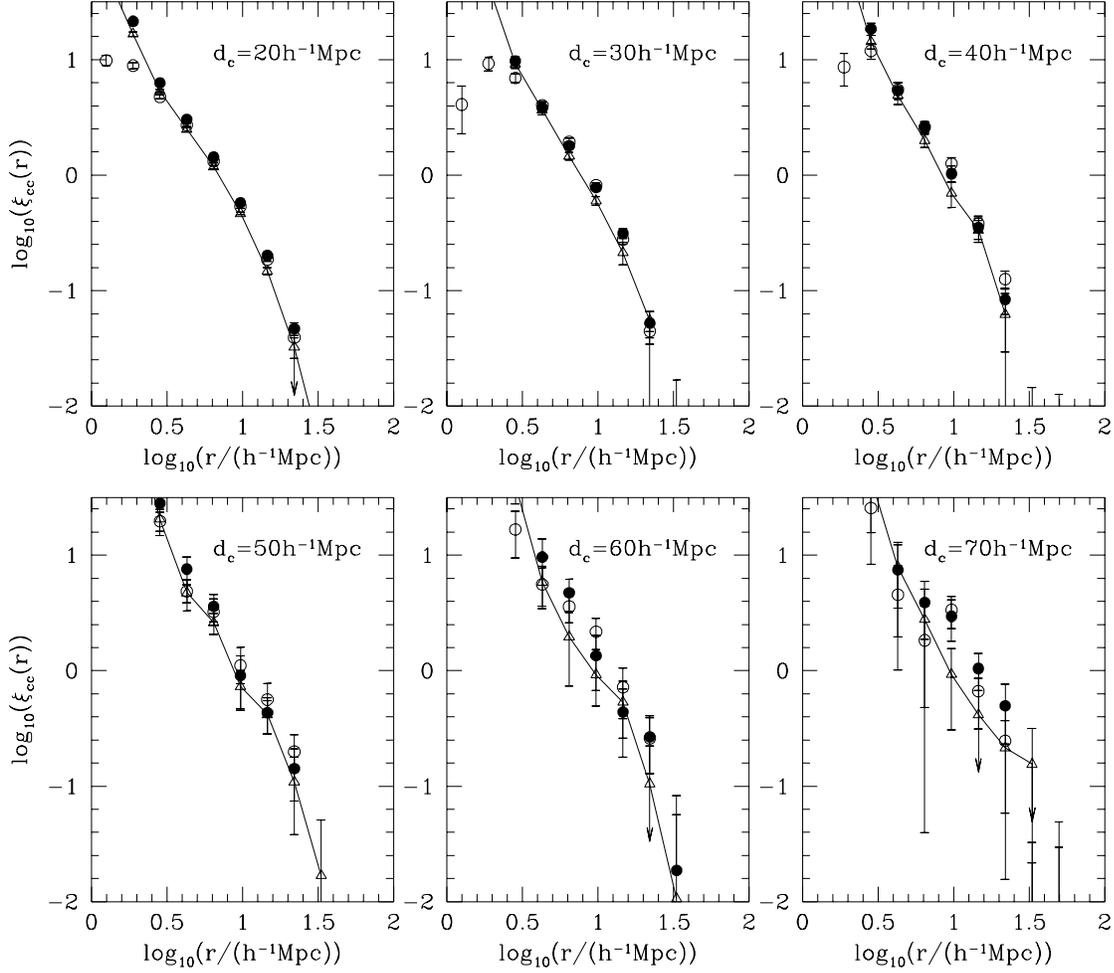}{441.0pt}}
\caption{Real space correlation functions for SO groups in a CDM model
with $\sigma_{8}=0.63$. The different symbols correspond to catalogues
selected according to mass (open circles), velocity dispersion (filled
circles), and ``X-ray'' luminosity (open triangles).  The lines in each
panel link the triangles and appear also on Fig.~4 for ease of
comparison.}
\end{figure*}

\begin{figure*}
\centering
\centerline{\epsfysize=15.5truecm \figinsert{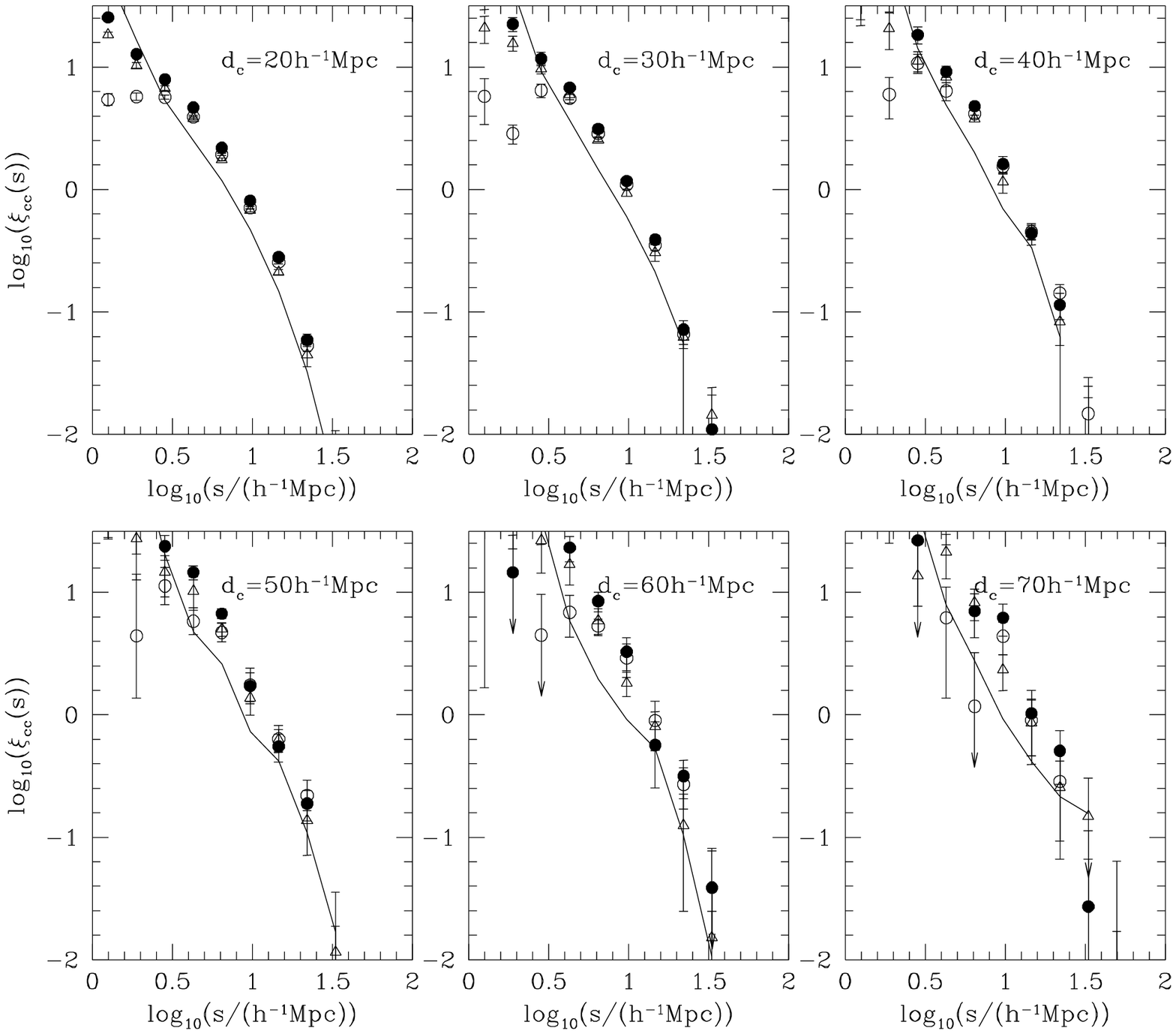}{441.0pt}}
\caption{Redshift space correlation functions for SO groups in a CDM model 
with $\sigma_{8}=0.63$. The different symbols correspond to catalogues
according to mass (open circles), velocity dispersion (filled circles), and
``X-ray'' luminosity (open triangles). The solid lines are the same as those
in Fig.~3 and are repeated here so that the increase in $\xi_{cc}$ caused
by using redshift distances rather than true distances may be seen clearly.}
\end{figure*}

In summary, we have used five different group-finders: FOF, SO, SOX and the
two employed by Croft \& Efstathiou (1994), CE(0.5) and CE(1.5). From the
FOF and SO groups, cluster catalogues selected according to mass, velocity
dispersion and X-ray luminosity were produced.  For the SOX group-finder,
clusters were selected either by X-ray luminosity or by velocity
dispersion. The cluster catalogues produced from the CE(0.5) and CE(1.5)
groups were selected according to mass only, in the same way as the clusters analyzed
by Croft \& Efstathiou (1994). In what follows, we indicate with a
subscript the statistic used for constructing a catalogue from the list
of groups obtained with a particular group-finder. Thus, for example,
${\rm FOF}_{M}$ denotes clusters identified with the friends-of-friends
group-finder and selected according to mass.

\subsection{Correlation function estimator}

Cluster correlation functions in real space, $\xi_{\rm cc}(r)$, and in 
redshift space, $\xi_{\rm cc}(s)$, were obtained for each of our cluster
catalogues. In the latter case, the simulations were projected along one
of the principal axes and the component of each cluster's peculiar velocity
along that axis was added to the Hubble velocity. We used the estimator
\begin{equation}
 \xi_{\rm cc}(x)={{N_{\rm p}} \over {n_{\rm c}^2VdV} }-1,
\end{equation}
where $x$ denotes either $r$ or $s$, $N_{\rm p}$ is the number of cluster
pairs with separation in a bin of volume $dV$ centred at $x$, $n_{\rm
c}$ is the number density of clusters in the catalogue, and $V$ is the
total volume. Estimates from all 10 simulations were averaged and the
scatter amongst them used to obtain the error in $\xi_{\rm cc}$.

Our correlation functions are not well fit by a single power-law over the
entire range of pair separations sampled in our models. Thus, to estimate the
correlation length, $x_0$, we made a two-parameter
$\chi^{2}$ fit of the form,
\begin{equation}
 \xi_{\rm cc}(x)=\left(\frac{x_{0}}{x}\right)^{\gamma}
\end{equation}
over a limited range in $x$, near the value where $\xi_{\rm cc}=1$. By 
fitting over a narrow range in $x$, our inferred values of $x_0$ 
do not depend strongly on the value of the slope, $\gamma$, but our 
estimates of $\gamma$ are only applicable in this limited range of 
pair separations. The actual range in $x$ used depends on the amplitude
of $\xi_{\rm cc}$, but is well approximated by the limits
\begin{equation}
 \frac{d_{\rm c}-10}{150}+0.5< {\rm log}_{10}(x)
< \frac{d_{\rm c}-10}{150}+1.0 ,
\end{equation}
where $d_{\rm c}$ and $x$ are measured in $\Mpc$.
This range corresponds to four or five radial bins around $x=x_{0}$.
The variances of the fit parameters were recovered from the diagonal
elements of the covariance matrix.

\section{Results}

Tables~1 and 2 give correlation lengths and power-law slopes for real
and redshift space correlation functions, for a selection of our cluster
catalogues. (Error bars were obtained as described in section 2.4). 
Full correlation functions are plotted in Figs.~3 and 4 for
samples constructed using a subset of our selection statistics and
spanning a wider range of values of $d_{\rm c}$ than those shown in the tables.
We now discuss several trends apparent in the data. We focus exclusively on
estimates of the correlation length, and ignore variations in the
slope, $\gamma$, since it is most often the former that is used to compare
models with the data. (For reference, we list our estimates of $\gamma$ in
the tables, recalling that they refer exclusively to the region where 
$\xi_{\rm cc}\simeq 1$.)

\begin{table*}
\centering
\begin{center}
\caption{The real space
correlation length, $r_0$, and the slope of the correlation function,
$\gamma$,  for a variety
of group-finders and cluster selection criteria for two different values of
mean cluster separation, $d_{\rm c}$, and normalization, $\sigma_8$. Errors
are $1\sigma$ and were estimated as described in section 2.4.}

\begin{tabular}{llllll} \hline
Group-Finder &
Selection &
\multicolumn{1}{l} {$d_{\rm c} = 30 \Mpc$} &
\multicolumn{1}{l} {$d_{\rm c} = 50 \Mpc$} &
\multicolumn{1}{l} {$d_{\rm c} = 30 \Mpc$} &
\multicolumn{1}{l} {$d_{\rm c} = 50 \Mpc$} \\
 & & $\sigma_8=0.5$ & $\sigma_8=0.5$ & $\sigma_8=0.63$ & $\sigma_8=0.63$ \\
\hline
 FOF & Mass & $r_0 = 8.50 \pm 0.15$ & $10.15 \pm 0.79$ &
$8.79 \pm 0.18$ & $12.01 \pm 0.58$ \\
  & & $\smash{\gamma}\hphantom{'} = \hphantom{gap}2.15 \pm 0.11$ & $\hphantom{gap}2.13 \pm 0.57$ & $\hphantom{gap}2.18 \pm 0.13$ &
$\hphantom{gap}2.63 \pm 0.37$ \\
 SO & Mass & $r_0 = 8.56 \pm 0.15$ & $11.24 \pm 0.70$ &
$8.73 \pm 0.17$ & $10.80 \pm 0.84$ \\
  & & $\smash{\gamma}\hphantom{'} = \hphantom{gap}2.18 \pm 0.11$ & $\hphantom{gap}2.25 \pm 0.49$ & $\hphantom{gap}2.22 \pm 0.11$ &
$\hphantom{gap}2.21 \pm 0.61$ \\
 CE(0.5) & Mass & $r_0 = 8.31 \pm 0.17$ & $7.97 \pm 0.71$ &
$8.29 \pm 0.17$ & $9.59 \pm 0.52$ \\
  & & $\smash{\gamma}\hphantom{'} = \hphantom{gap}2.02 \pm 0.12$ & $\hphantom{gap}1.89 \pm 0.48$ & $\hphantom{gap}2.14 \pm 0.12$ &
$\hphantom{gap}3.09 \pm 0.64$ \\
 CE(1.5) & Mass & $r_0 = 8.80 \pm 0.13$ & $10.70 \pm 0.64$ &
$8.84 \pm 0.16$ & $11.22 \pm 0.64$ \\
  & & $\smash{\gamma}\hphantom{'} = \hphantom{gap}2.27 \pm 0.09$ & $\hphantom{gap}2.53 \pm 0.51$ & $\hphantom{gap}2.42 \pm 0.12$ &
$\hphantom{gap}2.96 \pm 0.55$ \\
 FOF & $v$ & $r_0 = 8.64 \pm 0.16$ & $8.45 \pm 0.71$ &
$8.75 \pm 0.20$ & $10.12 \pm 0.69$ \\
  & & $\smash{\gamma}\hphantom{'} = \hphantom{gap}2.21 \pm 0.12$ & $\hphantom{gap}2.48 \pm 0.73$ & $\hphantom{gap}2.19 \pm 0.15$ &
$\hphantom{gap}2.74 \pm 0.66$ \\
 SO & $v$ & $r_0 = 8.54 \pm 0.14$ & $9.28 \pm 0.50$ &
$8.59 \pm 0.17$ & $9.89 \pm 0.68$ \\
  & & $\smash{\gamma}\hphantom{'} = \hphantom{gap}2.19 \pm 0.11$ & $\hphantom{gap}2.85 \pm 0.66$ & $\hphantom{gap}2.04 \pm 0.11$ &
$\hphantom{gap}2.67 \pm 0.65$ \\
 SOX & $v$ & $r_0 = 8.26 \pm 0.18$ & $9.00 \pm 0.76$ &
$8.32 \pm 0.19$ & $10.29 \pm 0.66$ \\
  & & $\smash{\gamma}\hphantom{'} = \hphantom{gap}1.95 \pm 0.12$ & $\hphantom{gap}1.91 \pm 0.46$ & $\hphantom{gap}2.17 \pm 0.15$ &
$\hphantom{gap}2.81 \pm 0.66$ \\
 FOF & $L_{\rm x}$ & $r_0 = 7.88 \pm 0.19$ & $9.14 \pm 0.73$ &
$7.87 \pm 0.16$ & $8.83 \pm 0.80$ \\
  & & $\smash{\gamma}\hphantom{'} = \hphantom{gap}2.33 \pm 0.17$ & $\hphantom{gap}2.60 \pm 0.98$ & $\hphantom{gap}2.11 \pm 0.14$ &
$\hphantom{gap}1.99 \pm 0.50$ \\
 SO & $L_{\rm x}$ & $r_0 = 7.34 \pm 0.19$ & $6.41 \pm 1.12$ &
$7.72 \pm 0.17$ & $9.46 \pm 0.62$ \\
  & & $\smash{\gamma}\hphantom{'} = \hphantom{gap}2.17 \pm 0.15$ & $\hphantom{gap}2.26 \pm 0.85$ & $\hphantom{gap}2.27 \pm 0.15$ &
$\hphantom{gap}2.72 \pm 0.77$ \\
 SOX & $L_{\rm x}$ & $r_0 = 7.76 \pm 0.16$ & $5.28 \pm 1.80$ &
$8.05 \pm 0.18$ & $10.02 \pm 0.70$ \\
  & & $\smash{\gamma}\hphantom{'} = \hphantom{gap}2.26 \pm 0.13$ & $\hphantom{gap}1.53 \pm 0.85$ & $\hphantom{gap}2.28 \pm 0.13$ &
$\hphantom{gap}2.45 \pm 0.60$ \\
\hline
\end{tabular}
\end{center}
\end{table*}

\begin{table*}
\centering
\begin{center}
\caption{The redshift space
correlation length, $s_0$, and the slope of the correlation function,
$\gamma$,  for a variety
of group-finders and cluster selection criteria for two different values of
mean cluster separation, $d_{\rm c}$, and normalization, $\sigma_8$. Errors
are $1\sigma$ and were estimated as described in section 2.4.}
\begin{tabular}{llllll} \hline
Group-Finder &
Selection &
\multicolumn{1}{l} {$d_{\rm c} = 30 \Mpc$} &
\multicolumn{1}{l} {$d_{\rm c} = 50 \Mpc$} &
\multicolumn{1}{l} {$d_{\rm c} = 30 \Mpc$} &
\multicolumn{1}{l} {$d_{\rm c} = 50 \Mpc$} \\
 & & $\sigma_8=0.5$ & $\sigma_8=0.5$ & $\sigma_8=0.63$ & $\sigma_8=0.63$ \\
\hline
 FOF & Mass & $s_0 = 9.31 \pm 0.16$ & $10.76 \pm 0.76$ &
$9.93 \pm 0.16$ & $12.33 \pm 0.65$ \\
  & & $\smash{\gamma}\hphantom{'} = \hphantom{gap}2.46 \pm 0.12$ & $\hphantom{gap}2.02 \pm 0.47$ & $\hphantom{gap}2.29 \pm 0.09$ &
$\hphantom{gap}2.61 \pm 0.35$ \\
 SO & Mass & $s_0 = 9.30 \pm 0.15$ & $11.82 \pm 0.62$ &
$9.81 \pm 0.16$ & $12.22 \pm 0.76$ \\
  & & $\smash{\gamma}\hphantom{'} = \hphantom{gap}2.55 \pm 0.10$ & $\hphantom{gap}2.59 \pm 0.47$ & $\hphantom{gap}2.36 \pm 0.08$ &
$\hphantom{gap}2.41 \pm 0.43$ \\
 CE(0.5) & Mass & $s_0 = 9.25 \pm 0.16$ & $10.45 \pm 0.53$ &
$9.89 \pm 0.16$ & $11.26 \pm 0.56$ \\
  & & $\smash{\gamma}\hphantom{'} = \hphantom{gap}2.30 \pm 0.10$ & $\hphantom{gap}2.58 \pm 0.44$ & $\hphantom{gap}2.04 \pm 0.09$ &
$\hphantom{gap}2.64 \pm 0.43$ \\
 CE(1.5) & Mass & $s_0 = 9.38 \pm 0.16$ & $10.89 \pm 0.56$ &
$9.89 \pm 0.15$ & $11.90 \pm 0.64$ \\
  & & $\smash{\gamma}\hphantom{'} = \hphantom{gap}2.52 \pm 0.09$ & $\hphantom{gap}2.70 \pm 0.47$ & $\hphantom{gap}2.41 \pm 0.10$ &
$\hphantom{gap}2.71 \pm 0.40$ \\
 FOF & $v$ & $s_0 = 9.31 \pm 0.14$ & $9.28 \pm 0.71$ &
$10.04 \pm 0.17$ & $11.07 \pm 0.66$ \\
  & & $\smash{\gamma}\hphantom{'} = \hphantom{gap}2.51 \pm 0.11$ & $\hphantom{gap}2.79 \pm 0.67$ & $\hphantom{gap}2.45 \pm 0.10$ &
$\hphantom{gap}3.12 \pm 0.62$ \\
 SO & $v$ & $s_0 = 9.59 \pm 0.16$ & $10.18 \pm 0.53$ &
$10.29 \pm 0.15$ & $12.03 \pm 0.58$ \\
  & & $\smash{\gamma}\hphantom{'} = \hphantom{gap}2.53 \pm 0.12$ & $\hphantom{gap}3.14 \pm 0.54$ & $\hphantom{gap}2.36 \pm 0.07$ &
$\hphantom{gap}3.05 \pm 0.34$ \\
 SOX & $v$ & $s_0 = 9.42 \pm 0.16$ & $11.61 \pm 0.60$ &
$10.32 \pm 0.15$ & $12.02 \pm 0.56$ \\
  & & $\smash{\gamma}\hphantom{'} = \hphantom{gap}2.52 \pm 0.13$ & $\hphantom{gap}2.74 \pm 0.34$ & $\hphantom{gap}2.41 \pm 0.08$ &
$\hphantom{gap}2.99 \pm 0.40$ \\
 FOF & $L_{\rm x}$ & $s_0 = 8.67 \pm 0.16$ & $9.39 \pm 0.94$ &
$9.19 \pm 0.15$ & $11.10 \pm 0.56$ \\
  & & $\smash{\gamma}\hphantom{'} = \hphantom{gap}2.60 \pm 0.12$ & $\hphantom{gap}2.00 \pm 0.70$ & $\hphantom{gap}2.46 \pm 0.10$ &
$\hphantom{gap}2.82 \pm 0.59$ \\
 SO & $L_{\rm x}$ & $s_0 = 8.53 \pm 0.14$ & $7.94 \pm 0.49$ &
$9.44 \pm 0.15$ & $11.65 \pm 0.62$ \\
  & & $\smash{\gamma}\hphantom{'} = \hphantom{gap}2.32 \pm 0.12$ & $\hphantom{gap}2.69 \pm 0.47$ & $\hphantom{gap}2.45 \pm 0.10$ &
$\hphantom{gap}2.74 \pm 0.33$ \\
 SOX & $L_{\rm x}$ & $s_0 = 8.72 \pm 0.14$ & $9.21 \pm 0.59$ &
$9.18 \pm 0.13$ & $10.98 \pm 0.69$ \\
  & & $\smash{\gamma}\hphantom{'} = \hphantom{gap}2.53 \pm 0.13$ & $\hphantom{gap}2.37 \pm 0.55$ & $\hphantom{gap}2.33 \pm 0.09$ &
$\hphantom{gap}2.64 \pm 0.55$ \\
\hline
\end{tabular}
\end{center}
\end{table*}

\subsection{Dependence of the real space correlation function
on cluster selection and abundance}

We first consider the effect on $\xcc$ of varying the procedure for
identifying and selecting clusters in real space. For each choice of
selection statistic ($M$, $v$ and $L_{\rm x}$), FOF and SO clusters
give consistent results in almost all cases. We conclude that for the
groups we are considering -- the most massive in the simulations --
the FOF and SO finders essentially pick out the same objects. SOX clusters
also tend to have similar values of $r_0$ as FOF and SO clusters,
whether they are selected by $v$ or $L_{\rm x}$.

Our real space results for CE clusters are in excellent agreement with
those obtained by Croft \& Efstathiou (1994). Note that whilst the ${\rm
CE(1.5)}_{M}$ clusters give results consistent with those of
FOF$_M$ and SO$_M$, the ${\rm CE(0.5)}_{M}$ clusters give the smallest
values of $r_0$ of any mass selected clusters. The largest difference
between CE(0.5)$_M$ and SO$_M$ clusters is 3.3 \hmpc and occurs for
$d_{\rm c}=50$\hmpc and $\sigma_8=0.5$.

For any given cluster finding algorithm there are often trends either with
the selection statistic or with $\sigma_8$. For example, when $d_{\rm c}$ is
large, clusters selected by mass have larger values of $r_0$ than clusters
selected by $v$, with a largest difference of 2.8 \hmpc between SO$_M$ and
FOF$_{v}$ clusters for $\sigma_8=0.5$. Clusters selected by X-ray
luminosity tend to be slightly more weakly clustered than clusters
selected by mass or velocity dispersion. This is because of the scatter in
the $L_{\rm x}-T_{\rm x}$ relation and the trend of clustering strength
with velocity dispersion.
Thus, for fixed $d_{\rm c}$ and
$\sigma_8$, the SO$_{L_{\rm x}}$ and SOX clusters tend to give smaller
values of $r_0$ than the other catalogues.
At $d_{\rm c}=30$\hmpc, where the statistical errors are small, the differences
between the correlation lengths measured from the various cluster catalogues
can exceed $5\sigma$ and at $d_{\rm c}=50$\hmpc the 
minimum and maximum are separated by at least $4\sigma$.
Our data also show a weak but significant trend for $r_0$
to increase with increasing $\sigma_8$. Typically, with the same
cluster definition at $\sigma_8=0.5$ and $\sigma_8=0.63$, the
differences in $r_0$ are $1$\hmpc or less.

A clear trend in our real space data is the tendency for the correlation
length to increase with increasing mean cluster separation. This trend is
stronger for the larger value of $\sigma_8$. Some illustrative
cases are plotted in Fig.~5 (for $v$-selected clusters). The data are
reasonably well fit by a linear relation, although for $d_{\rm c}>40$\hmpc the
uncertainties are too large to rule out a flatter trend as advocated by
Croft \& Efstathiou (1994). A more detailed comparison with this and other
work is made in the next section.  For our SO$_{v}$ clusters, the 
$r_0-d_{\rm c}$
relation can be approximately fit by a linear relation of the form:

 \begin{equation}
 r_0 = (0.056 \pm 0.014)d_{\rm c} + (6.36 \pm 0.69) \, \Mpc
 \end{equation}
and
 \begin{equation}
 r_0 = (0.090 \pm 0.009)d_{\rm c} + (5.82 \pm 0.44) \, \Mpc
 \end{equation}
for $\sigma_8 = 0.5$ and $0.63$ respectively.

In summary, the real space correlation length of rich clusters identified
in three dimensions is only weakly dependent 
on the normalization of the power spectrum,
but it can vary considerably depending on the procedure used to define
a cluster catalogue and on the abundance of the objects
under consideration. This variation can be much 
larger than the statistical uncertainties in the individual
determinations. For example, for $\sigma_8=0.5$ and $d_{\rm c}=30$ \hmpc, the
largest variation seen in Table~1 is 1.46 \hmpc,
compared with a typical uncertainty of $\sim 0.16$\hmpc in the individual
determinations.

\begin{figure}
\centering
\centerline{\epsfysize=9.5truecm \figinsert{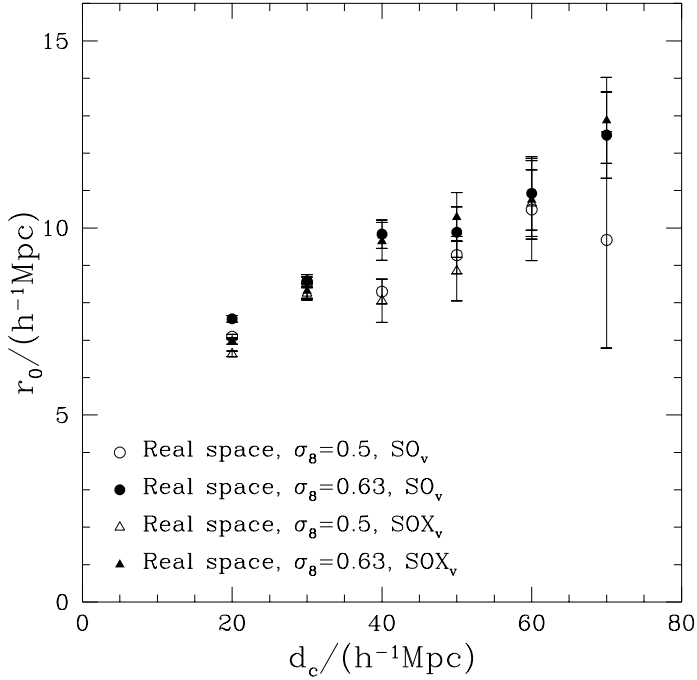}{441.0pt}}
\caption{Variation of the real space correlation length, $r_0$, with mean
intercluster separation, $d_{\rm c}$ for a selection of cluster catalogues.
Real space correlation functions were calculated for catalogues
selected according to 1-D velocity dispersion, $v$, for SO (circles) and
SOX (triangles) clusters. Open symbols are for
$\sigma_{8}=0.5$ and filled symbols for~$0.63$.}
\end{figure}

\begin{figure}
\centering
\centerline{\epsfysize=9.5truecm \figinsert{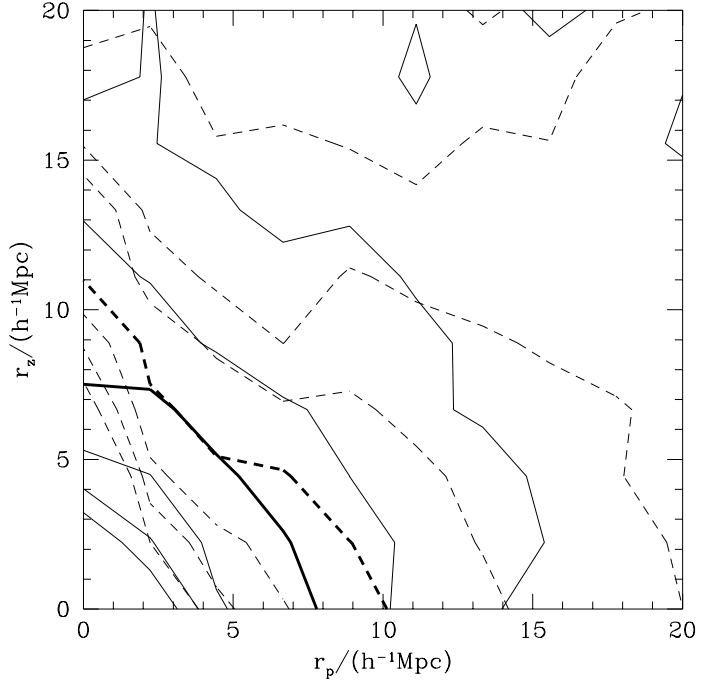}{441.0pt}}
\caption{Contours of constant $\xi_{\rm cc}$ for a CDM model with
$\sigma_{8}=0.63$. Results are shown for SO$_v$ clusters with $d_{\rm
c}=20 \Mpc$. The levels are at $0,0.2,0.5,1,2,3$ and $4$. Real space
correlations are plotted as full lines and redshift space correlations as
dashed lines, with the $\xi_{\rm cc}=1$ contour in bold.} 
\end{figure}

\begin{figure}
\centering
\centerline{\epsfysize=9.5truecm \figinsert{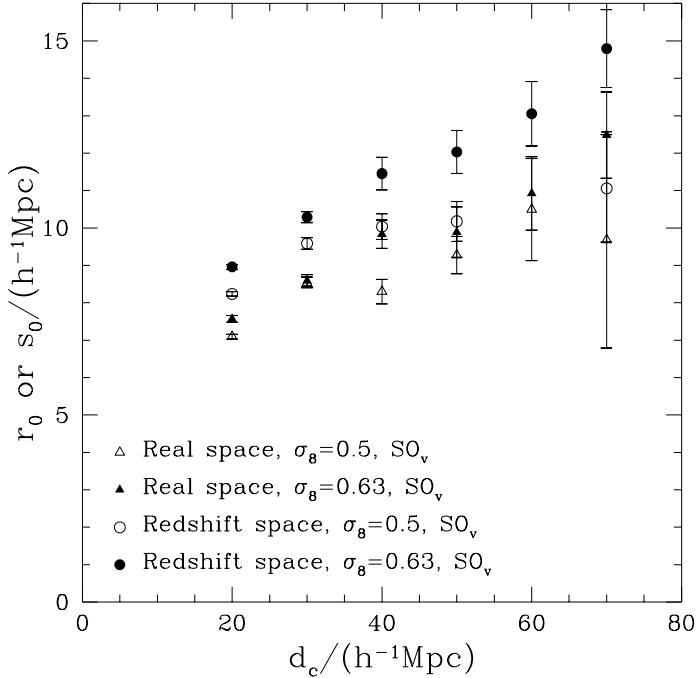}{441.0pt}}
\caption{Variation of the redshift space correlation length, $s_0$, with
mean intercluster separation, $d_{\rm c}$, 
for a selection of cluster catalogues.
Data are for SO groups selected according to velocity dispersion.  Open
symbols show results for $\sigma_{8}=0.5$ and filled symbols for $\sigma_8=
0.63$. Circles represent our redshift space results and triangles the
corresponding real space results from Fig.~5 which are included for
comparison.}
\end{figure}

\subsection{Redshift space effects}

When redshift distances rather than true distances are used, the correlation
function is distorted in various ways. Structures on small scales are
smeared out by peculiar velocities, while structures on large scales are
amplified by coherent infall (Kaiser 1987). As a result, the correlation
function in redshift space is flatter on small scales, steeper on
intermediate scales and has a larger amplitude on large scales than the
real space correlation function. These effects are readily apparent in
Fig.~6 which shows contour plots of $\xi_{\rm cc}$ as a function of
projected separation on the sky, $r_{p}$, and distance along the
line-of-sight, $r_{z}$. For small values of $r_p$, the contours of constant
$\xi_{\rm cc}$ 
are elongated along the $r_z$ direction because of smearing and for
large values of $r_p$ they are elongated along the $r_p$ direction as a result
of coherent infall.

A selection of our redshift space correlation functions are plotted and
compared to their real space counterparts in Fig.~4. In all cases, the
net effect of redshift space distortions is to increase the amplitude of
the correlation function on scales $3 \lsim s/(\Mpc) \lsim 30$. As a result,
the values of $s_0$ (the redshift space separation at which
$\xi_{\rm cc}(s)=1$) are significantly larger than the corresponding values of
$r_0$. As expected, the differences are greater for larger values of
$\sigma_8$ since the induced peculiar velocities grow with the amplitude of
the mass fluctuations. These effects are further demonstrated in Tables~1
and 2. The enhancement of the correlation length in redshift space depends
somewhat on the group-finder and selection statistic used. On average, $s_0$
is larger than $r_0$ by $1.1$\hmpc for $\sigma_8=0.5$ and by $1.4$\hmpc for
$\sigma_8=0.63$. Note, however, that in individual cases the redshift space
enhancement can be considerably larger than this.

The increase in the correlation length with increasing intercluster
separation is slightly 
more pronounced in redshift space than in real space. An
illustrative case, SO$_{v}$ clusters, is shown and compared with the 
corresponding real space data in Fig.~7. For $\sigma_8=0.63$,
$s_0$ grows approximately linearly with $d_{\rm c}$ 
out to the largest values of
$d_{\rm c}$ considered, $d_{\rm c}=70$\hmpc, at which $s_0 \simeq 15$\hmpc. The
relation between $s_0$ and $d_{\rm c}$ is approximately given by
\begin{equation}
 s_0 = (0.063 \pm 0.020) d_{\rm c} + (7.31 \pm 0.72) \, \Mpc
\end{equation}
and
\begin{equation}
 s_0 = (0.109 \pm 0.007) d_{\rm c} + (6.87 \pm 0.35) \, \Mpc
\end{equation}
for $\sigma_8 = 0.5$ and $0.63$ respectively.

\begin{figure}
\centering
\centerline{\epsfysize=9.5truecm \figinsert{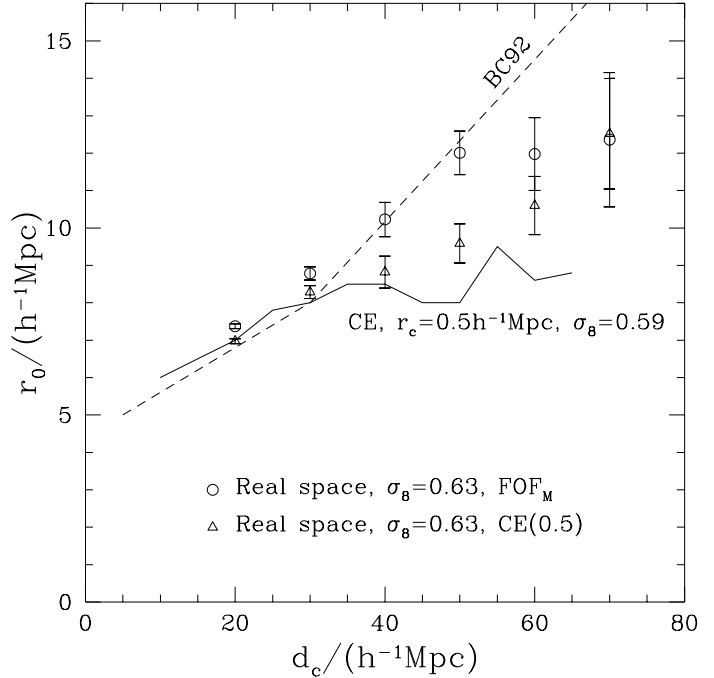}{441.0pt}}
\caption{Comparison of simulation results from different studies. The plot
shows the real space correlation length, $r_0$, as a function of the mean
intercluster separation, $d_{\rm c}$. The dashed line gives the results of
Bahcall \& Cen (1992) which should be compared with our results for FOF$_M$
clusters (circles). The solid line gives results from Croft \& Efstathiou
(1994) which should be compared with our results for CE(0.5) (triangles). 
Within the statistical errors, our results agree with the two
other studies even though these are inconsistent with one another.} 
\end{figure}

\begin{figure}
\centering
\centerline{\epsfysize=9.5truecm \figinsert{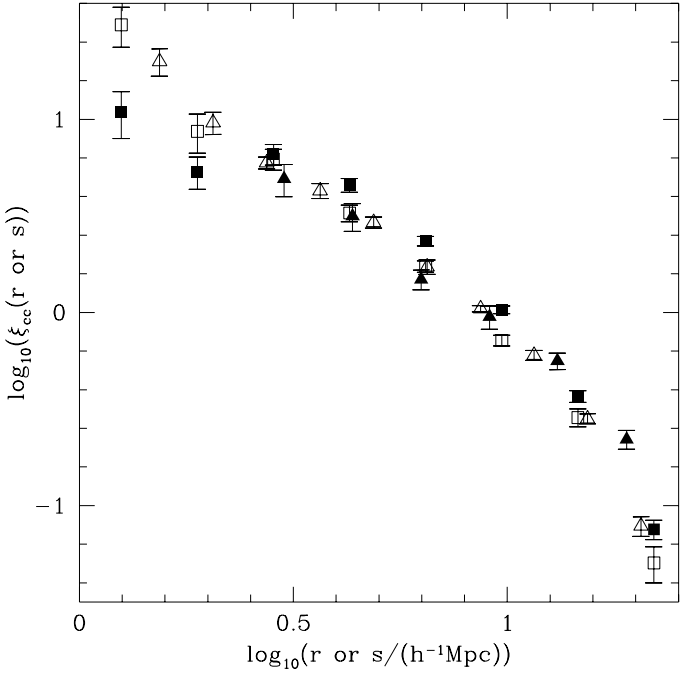}{441.0pt}}
\caption{Comparison with the results of Croft \& Efstathiou
(1994). Correlation functions from the present study are shown by squares
and those from Croft \& Efstathiou by triangles. Open symbols show data in
real space and filled symbols in redshift space.  All correlation functions
are for CE(0.5) groups with $d_{\rm c}=30 \Mpc$, but our simulations have
$\sigma_{8}=0.63$ whereas those of Croft \& Efstathiou have
$\sigma_8=1.0$.}
\end{figure}

\begin{figure}
\centering
\centerline{\epsfysize=15.0truecm \figinsert{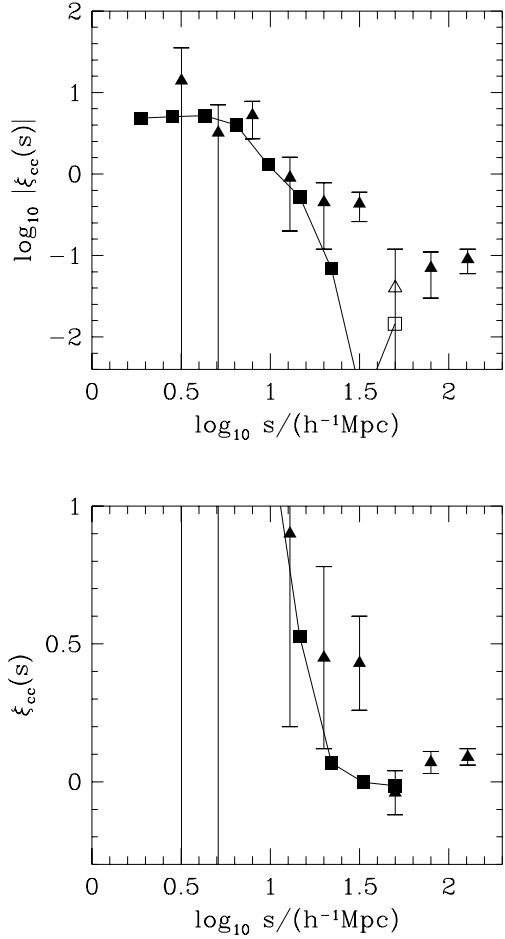}{441.0pt}}
\caption{Comparison between our CDM cluster correlation function and data
for X-ray selected clusters. The squares show our redshift space results
for SOX$_{L_{\rm x}}$ clusters with $d_{\rm c}=40 \Mpc$ and
$\sigma_{8}=0.63$. The triangles show the correlation function estimated by
Romer \etal (1994) for a sample of clusters selected using a combination of
ROSAT X-ray data and optical data.  The upper panel shows $\log_{10} \vert
\xi_{\rm c}(s)\vert$, with open symbols corresponding to values of $\xi_{\rm
cc}(s)<0$. The lower panel is a linear-log plot of the region where
$\xi_{\rm cc}<1$ which shows the large scale behaviour more clearly.}
\end{figure}

\begin{figure}
\centering
\centerline{\epsfysize=15.0truecm \figinsert{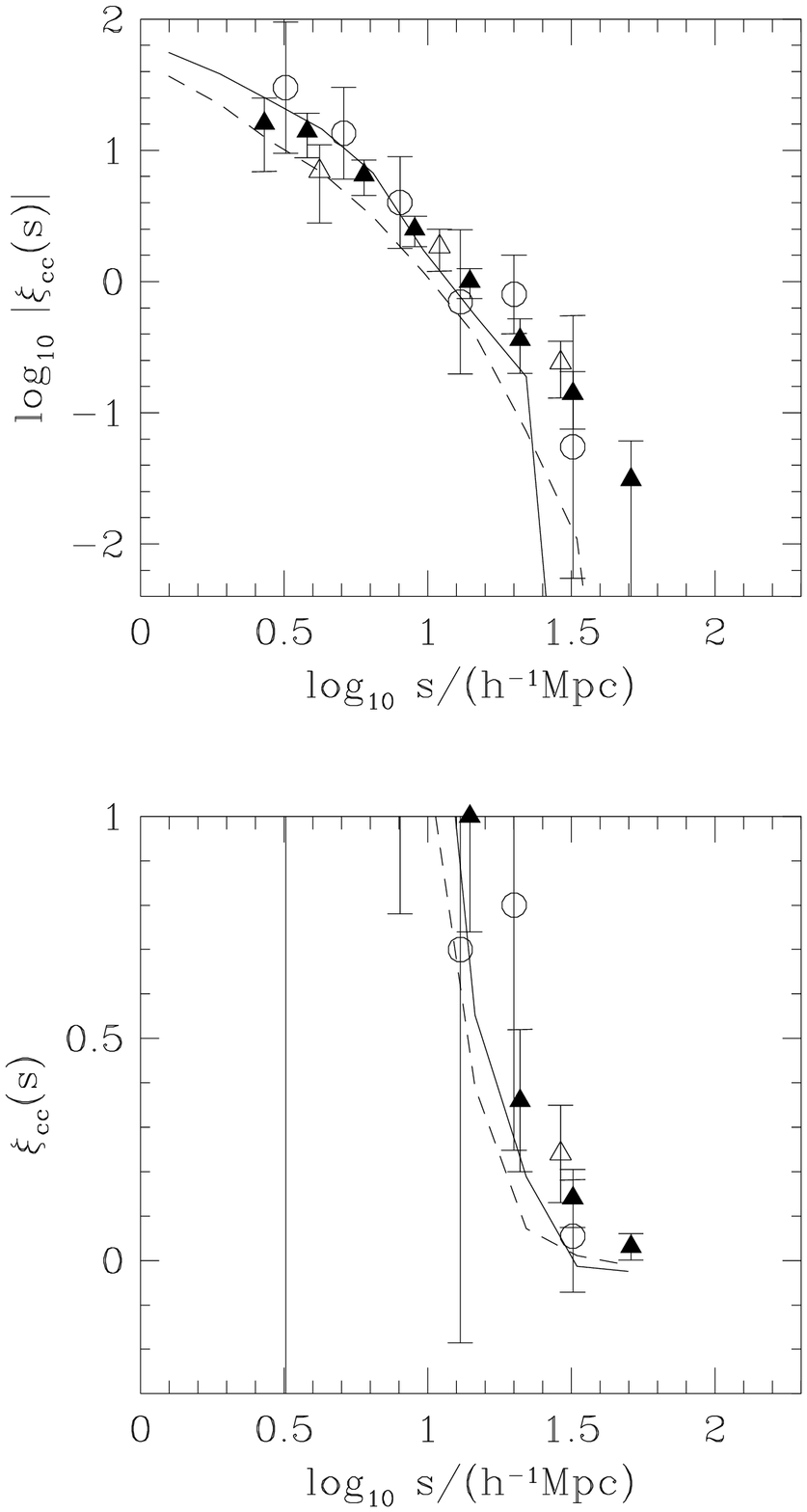}{441.0pt}}
\caption{Comparison between CDM cluster correlation functions and data for
optically selected clusters. The triangles show correlation functions for
APM clusters: filled triangles for the $d_{\rm c}=31$\hmpc sample of Dalton
\etal (1994), and open triangles for the $d_{\rm c}=45$\hmpc sample of
Efstathiou \etal (1992). The open circles show the correlation function
estimated by Nichol \etal (1992) 
for the $d_{\rm c}\simeq 50$ \hmpc clusters in the 
Durham-Edinburgh 
survey. The dashed line shows our model results for SO$_v$ clusters, with
$d_{\rm c}=30 \Mpc$, and the solid line for SO$_v$ clusters with $d_{\rm
c}=50 \Mpc$. All model correlation functions were calculated in redshift
space.}
\end{figure}

\section{COMPARISON WITH PREVIOUS SIMULATIONS} 

Cluster correlation functions calculated from N-body simulations of the
CDM model have been published previously by White \etal (1987), Bahcall \&
Cen (1992), and Croft \& Efstathiou (1994). The first of these studies
sampled a relatively small volume so the resulting correlation functions
have large uncertainties; they are consistent with our results.

Bahcall \& Cen (1992) simulated a single cube of side $400$\hmpc, in the
standard CDM model, and calculated cluster correlation functions in real
space only (Cen, private communication). They identified groups of
particles using a variant of our FOF algorithm and selected clusters
according to mass. Thus, for a fair comparison, we consider our
catalogue of FOF$_M$ clusters. The comparison is done in Fig.~8 where we
plot correlation lengths in real space as a function of mean intercluster
separation. For small values of $d_{\rm c}$ (corresponding to relatively poor
clusters), Bahcall \& Cen's results lie only slightly below ours and for
intermediate values the agreement is very good. For the rarest objects
($d_{\rm c}\gsim 50$\hmpc), 
Bahcall \& Cen apparently find that the $r_0$-$d_{\rm c}$
relation continues to increase linearly whereas our relation flattens off.
Unfortunately, Bahcall \& Cen do not quote uncertainties in their
estimates but, since the volume they simulated is only about 40\% of the
volume we have simulated, their error bars will typically be $\sim 1.6$
larger than ours and so the disagreement at large values of $d_{\rm c}$ is only
marginally significant. 

In Fig.~8 we also compare our real space results with those of Croft \&
Efstathiou (1994). Like Bahcall \& Cen, they examined only one cluster
selection algorithm (CE in Section~2) which we deliberately included in our
list for comparison purposes. Again, the agreement is very good,
particularly for $d_{\rm c} \sim 40$\hmpc, where the error bars are small. For
larger values of $d_{\rm c}$ the results are still statistically consistent
although Croft \& Efstathiou's values lie somewhat below ours (Croft \&
Efstathiou do give error bars but, for clarity, we have omitted these in
the figure; they are slightly smaller than ours.) As pointed out by
Croft \& Efstathiou, their results disagree with those of Bahcall \& Cen,
except for the smallest 
values of $d_{\rm c}$. Croft \& Efstathiou suggested that
this discrepancy could be due to a statistical fluctuation in Bahcall \&
Cen's single simulation. 
The two simulations also differ in their value of $\sigma_8$ ($0.77$ in the
case of Bahcall \& Cen; $0.59$ in the case of Croft \& Efstathiou; and an
intermediate value in our case). However, as Fig. 7 shows, the effect of
$\sigma_8$ on the real space correlation length is too weak to account for 
the difference between the results of Bahcall \& Cen and those of
Croft \& Efstathiou.
Fig.~8 suggests that the discrepancy might be
caused by the use of different cluster selection criteria. When
we use similar selection criteria, we find reasonably good 
agreement with both studies. 

The previous comparison referred exclusively to correlation lengths in real
space. Bahcall \& Cen do not give any results in redshift space, but Croft
\& Efstathiou do and a comparison of their results and ours (for the same
cluster selection criteria) is made in Fig.~9. Here we plot the full
correlation functions for 
CE(0.5) clusters with $d_{\rm c}=30$\hmpc, both in real
space (open symbols) and in redshift space (filled symbols). The agreement
in real space is excellent, confirming our earlier conclusion from
Fig.~8. In redshift space, on the other hand, there are some
discrepancies, particularly at small and intermediate pair separations,
where our correlation function lies systematically above Croft \&
Efstathiou's. The differences are small but significant given the small
quoted errors. For example, at $s=6.3$\hmpc, our value of $\xi_{\rm cc}$ is
about $60 \%$ higher than Croft \& Efstathiou's. These differences are
likely to be an underestimate since our simulation has $\sigma_8=0.63$,
whereas theirs has $\sigma_8=1$ and, as we have seen, in redshift
space $\xi_{\rm cc}(s)$ is larger for larger values of $\sigma_8$. Apart from
this difference in $\sigma_8$, the only other difference between the two
analyses is the sampling strategy in the computation of $\xi_{\rm cc}(s)$. Our
estimate is based on a straightforward computation using all the clusters
in the simulations satisfying the selection criteria, whereas Croft \&
Efstathiou averaged over several realizations of subcatalogues with the
same abundance and selection function as the APM cluster catalogue of
Dalton \etal (1992).  Provided the correct selection function is used
this procedure will give an unbiased estimate of the true
correlation function in the simulation as a whole -- the quantity which we
have calculated directly. Possible explanations for the relatively small
discrepancies in Fig.~9 are residual systematic differences resulting from
different simulation techniques and the choice of $\sigma_8$.

In summary, apart from the small differences in redshift space just mentioned,
the cluster correlation functions in our study agree well with previous
published work, provided the comparison is made {\it for cluster catalogues
identified and selected in similar ways.} The apparent disagreement between
the work of Bahcall \& Cen (1992) and that of Croft \& Efstathiou (1994)
appears to have been largely caused by a different choice of group-finding
algorithm. 

\section{DISCUSSION AND CONCLUSIONS} 

Bahcall \& Cen (1992) and Croft \& Efstathiou (1994) compared their
simulation results to real data and in both cases concluded that the
correlation function of clusters is incompatible with the standard CDM
model, but is consistent with a low density CDM model with $\Omega h \simeq
0.2$. At first sight this consensus seems rather surprising since, as
Fig.~8 shows, the predicted cluster correlation functions in these two
studies disagree. The explanation is simply that the comparison in each
case was made against different datasets. Croft \& Efstathiou compared
their models to the APM cluster catalogue of Dalton \etal (1992) whereas
Bahcall \& Cen compared theirs to this catalogue and to Abell's catalogue
as well. As may be seen from Fig.~3 of Bahcall \& Cen (1992), their
cluster correlation function for the low density model does not agree
particularly well with the APM data and, as can be seen from Fig.~4 of
Croft \& Efstathiou (1994), their low density model strongly disagrees
with the Abell cluster data. Thus, the two studies were able to arrive at
the same conclusion because they compared different theoretical
predictions for the same cosmological models against datasets that exhibit
different clustering properties. 

We have found here that even in the idealized case where clusters are
identified in the three-dimensional mass distribution of a simulation,
significantly different outcomes for the cluster correlation function are
possible depending on how exactly the clusters are defined and on how the
data are analyzed. It is unclear which, if any, of the various possible
definitions of clusters in the simulations is appropriate for a comparison
with the real data. This difficulty is particularly severe in the case of
optical catalogues since the identification of clusters in the projected
galaxy distribution is very different from the identification of clusters
in the three-dimensional mass distribution of a simulation. Biases in
$\xcc$ arising from projection effects have been shown to be present in
Abell's catalogue (Sutherland 1988; Efstathiou \& Sutherland 1991; Dekel
\etal 1989). The lack of large anisotropies in $\xcc$ for APM clusters
suggests that this catalogue is largely unaffected by biases of this kind,
but this important feature by itself does not remove the ambiguity
regarding the identification of galaxy clusters in real catalogues with
mass clusters in simulations. Identifying cluster populations in both by
matching the spatial number density is not a unique procedure since, as we
have shown, even at a fixed number density, the cluster correlation
function depends, for example, on the statistic used to rank the clusters.
In practice, it seems likely that even larger uncertainties will be
introduced by the difficulty of determining the richness of clusters in
projection and by the associated uncertainties in the estimation of their
spatial number density. X-ray selected clusters provide, in principle,
cleaner observational samples, but even in this case the comparison with
theoretical models is restricted by the lack of reliable predictions.

Figs.~10 and 11 illustrate how some of the uncertainties we have
mentioned can affect the confidence with which a specific model is
constrained by measurements of $\xcc$. Here we compare predictions based on
the standard cold dark matter model -- the model which Bahcall \& Cen
(1992), Croft \& Efstathiou (1994) and Dalton \etal (1994) claim to be
strongly excluded by the cluster correlation function data -- with various
observational determinations of $\xcc$. In Fig.~10 we compare our
estimates for SOX ``X-ray selected'' clusters with the ROSAT data of Romer
\etal (1994). This catalogue is not volume-limited and thus contains
clusters with a range of intrinsic X-ray luminosities.  The simulations,
however, indicate that the variation of the correlation length of SOX
``X-ray selected'' clusters with cluster X-ray luminosity (or richness) is
small compared with the uncertainties in the data. Within these large
statistical errors, the agreement is good except, perhaps, on the largest
scales where the observed signal is small and could be affected by
systematic uncertainties in the mean number density of clusters. The size
of the discrepancy on large scales may be better appreciated in the
linear-log plot in the lower panel of this figure.

In Fig.~11, we compare our estimates for SO$_v$ clusters
(in redshift space) with data for ``optical'' clusters from the APM
(Efstathiou \etal 1992, Dalton \etal 1992, 1994) and EDCC (Nichol \etal
1992) catalogues. The APM sample shown by filled triangles has a mean
intercluster separation, $d_{\rm c}\simeq 31$\hmpc, and should be compared with
the dashed line which shows our model predictions for the same mean
intercluster separation.  The APM sample shown by open triangles has
$d_{\rm c}\simeq 45$\hmpc 
and the EDCC sample has $d_{\rm c}\simeq 50$\hmpc. These
should be compared with the solid line which shows our model results for
$d_{\rm c}=50$\hmpc. 
On intermediate scales only the denser sample (which has the
smallest error bars) is inconsistent with the model, but the discrepancy is
quite small and certainly much smaller than the discrepancy found by Dalton
\etal (1994) for the same model (cf their Fig.~4). The reason for this
difference is simply that the group-finder applied to the simulations by
Dalton \etal happens to give one of the lowest correlation functions of all the
group-finders that we have explored in this paper. On large scales there is
an indication that the data are more clustered than the models and, again,
the linear-log plot clearly shows that this discrepancy is 
small.

To summarize, large N-body simulations allow very precise estimates of the
cluster correlation function once a specific prescription for identifying
clusters is adopted. For a given cosmological model, the statistical
uncertainties in these predictions are small compared to the observational
errors for existing cluster samples. Unfortunately, they are also small
compared to the systematic variations exhibited by cluster samples
constructed from the simulations by making different assumptions.  Our
analysis shows that the exact form and amplitude of the correlation
function of clusters identified in the mass distribution of N-body
simulations depends on various factors. In rough order of importance these
include: (i) the group-finding algorithm and the statistic used to rank
clusters in a catalogue (eg. mass, velocity dispersion, ``X-ray''
luminosity, etc); (ii) the mean density of clusters in the catalogue; (iii)
whether clustering is measured in real or in redshift space; and (iv) the
assumed value of $\sigma_8$. These various factors can produce large
variations in the resulting clustering length. For example, in the range
most relevant to observational data, $30\le d_{\rm c} \le 60 \Mpc$, our
``X-ray selected" catalogues in a model with $\sigma_8=0.5$ 
have real space clustering lengths varying
between 7 and 10\hmpc, whereas catalogues selected by velocity dispersion
in a model with $\sigma_8=0.63$ have redshift space correlation lengths
varying from $10$ to $13 \Mpc$.  

Of the four complicating factors listed above, only the first two refer
directly to cluster catalogues. The third one should be straightforward to
eliminate but it has sometimes been ignored in comparisons of model
predictions with data (eg. Bahcall \& Cen 1992). Similarly, the value of
$\sigma_8$ appropriate to a given cosmological model is usually fixed from
other considerations such as the amplitude of fluctuations in the
temperature of the cosmic microwave background or the abundance of galaxy
clusters (White, Efstathiou \& Frenk 1993). There are, however,
uncertainties associated with this procedure arising, for example, from
possible contamination of the microwave background signal by tensor modes or
uncertainties in the masses of clusters.

The resolution of the cluster clustering debate will require further
observational and theoretical work. From the observational point of view,
progress will come from the analysis of large homogeneous samples of
clusters selected entirely from X-ray data or from large redshift surveys
such as the forthcoming SDSS and 2df galaxy surveys. From the theoretical
point of view, it will be necessary to model in detail the selection
procedures employed by observers. Artificial catalogues constructed from
cosmological simulations are a valuable aid, but several complications need
to be borne in mind. For example, to model the selection of APM clusters,
it is necessary to simulate the entire APM galaxy survey and this, in turn,
requires modelling the uncertain connection between the distribution of
dark matter and the formation sites of galaxies. Modelling cluster
catalogues constructed from X-ray data is a simpler problem theoretically
since it bypasses the complications associated with galaxy
formation. Nevertheless, it requires a better understanding of the
mechanisms that determine the total cluster X-ray luminosity than is
available at present. It is our view that until the theoretical predictions
are brought onto the observational plane, measurements of the cluster
correlation function cannot confidently be used to choose amongst competing
cosmological models.

\section*{Acknowledgements}
We thank Hugh Couchman for providing a copy of his AP$^3$M N-body code and
for giving valuable advice and support.  VRE acknowledges the support of a
PPARC studentship and SMC a PPARC Advanced
Fellowship. This work was supported in part by a PPARC rolling grant.

\end{document}